\documentclass[]{aa} 
\usepackage{epsfig,graphicx,lscape}
\usepackage{txfonts}
\begin{document}

   \title{The large- and small-scale Ca\,II\,K structure of the Milky Way from observations of 
          Galactic and Magellanic sightlines}

   \author{J.~V. Smoker
          \inst{1}\fnmsep\thanks{Based on observations taken from the ESO archive, programme IDs 
078.C-0493(A) and 171.D-0237(A).}
          \and
	  F.~P. Keenan \inst{2}
          \and
          A.~J. Fox \inst{3}
          }

   \offprints{J.~V. Smoker}

   \institute{European Southern Observatory, Alonso de Cordova 3107, 
              Vitacura, Santiago, Chile \\
              \email{jsmoker@eso.org} 
   \and Astrophysics Research Centre, School of Mathematics and Physics,
              Queen's University Belfast, Belfast, BT7 1NN, U.K. 
   \and Space Telescope Science Institute, 3700 San Martin Drive, Baltimore, MD 21218, U.S.A. \\
             }

   \date{Received Accepted}

 
  \abstract
   {}
   {By utilising spectra of early-type stellar probes of known distances in the same region of the sky, the large and small-scale (pc) structure of the 
    Galactic interstellar medium can be investigated. This paper 
    determines the variation in line strength of Ca\,{\sc ii} at 3933.661\AA \, as a function of probe separation for a large sample of stars, 
    including a number of sightlines in the Magellanic Clouds.}
   {FLAMES-GIRAFFE data taken with the Very Large Telescope towards early-type stars in 3 Galactic and 4 Magellanic open 
   clusters in Ca\,{\sc ii} are used to obtain the velocity, equivalent width, 
   column density and line width of interstellar Galactic calcium for a total of 657 stars, of which 
   443 are  
   Magellanic Cloud sightlines. In each cluster there are between 43-110 stars observed. Additionally, FEROS and UVES Ca\,{\sc ii} K 
   and Na\,{\sc i} D spectra of 21 Galactic 
   and 154 Magellanic early-type stars are presented and combined with data from the literature 
   to study the calcium column density -- parallax relationship.  }
   {For the four Magellanic clusters studied with FLAMES, the strength of the Galactic interstellar Ca\,{\sc ii} K equivalent width over transverse 
    scales from $\sim$0.05--9 pc is found to vary by factors of 
    $\sim$1.8--3.0, corresponding to column density variations of $\sim$0.3--0.5 dex in 
    the optically-thin approximation. 
    Using FLAMES, FEROS and UVES archive spectra, the minimum and maximum reduced equivalent width for 
    Milky Way gas is found to lie in the range $\sim$35--125 m\AA \, and $\sim$30--160 m\AA \, for Ca\,{\sc ii} K and Na\,{\sc i} D, respectively. 
    The range is consistent with a simple model of the ISM published by van Loon et al. (2009) 
    consisting of spherical cloudlets of filling factor $\sim$0.3, although other geometries are not ruled out. 
    Finally, the derived functional form for parallax ($\pi$) and Ca\,{\sc ii} column density ($N_{\rm CaII}$)
    is found to be $\pi$(mas)=1/(2.39$\times$10$^{-13}\times N_{\rm CaII}$(cm$^{-2}$) + 0.11). Our derived parallax is $\sim$25 per cent lower than 
    predicted by Megier et al. (2009) at a distance of $\sim$100 pc and $\sim$15 percent lower at a distance of $\sim$200 pc, 
    reflecting inhomogeneity in the Ca\,{\sc ii} distribution in the different sightlines studied.

    The full version including online material is available via the Astronomy and Astrophysics website http://www.aanda.org/articles/aa/olm/2015/10/aa25190-14/aa25190-14.html
}

   \titlerunning{Ca\,{\sc ii} K interstellar observations of 7 Galactic and MC open clusters}
   \maketitle
%

\section{Introduction}


\begin{figure*}
\setcounter{figure}{0}
\centering
\includegraphics{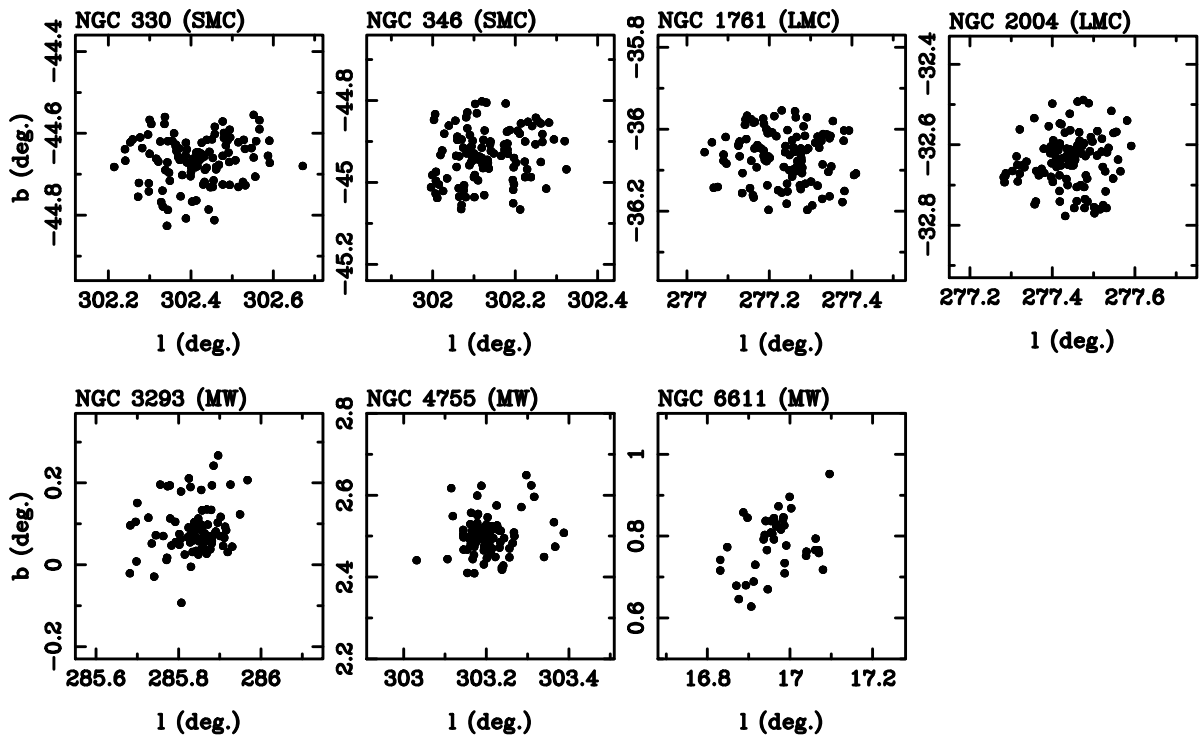}
\caption{FLAMES-GIRAFFE sample: Galactic coordinates of the sightlines for which low-velocity interstellar Ca\,{\sc ii} K was detected.}
\label{f_FLAMES_Star_lb}
\end{figure*}

Since its discovery in the interstellar medium more than 100 years ago (Hartmann 1904), the Ca\,{\sc ii} K line has been 
extensively studied. At 3933\AA \, it lies in a region of the spectrum free of strong telluric features and 
where CCD detectors are sensitive. Perhaps most importantly the transition itself is strong and hence 
easily detected in most high-resolution stellar spectra of medium signal to noise (S/N) ratio. Recent studies of Ca\,{\sc ii} include 
those by Albert et al (1993), Sembach et al. (1993), Welty et al. (1996; at very high 
spectral resolution to obtain detailed information on cloud velocity components), Wakker \& Mathis (2000; 
to determine the spread in the Ca/H ratio as a function of H\,{\sc i} column density), and Smoker et al (2003; to determine 
the Galactic scale-height of the transition). More recently, Megier et al. (2005, 2009) compared the strength of the 
Ca\,{\sc ii} K interstellar absorption with distances from the Hipparcos catalogue (ESA 1997) and open cluster 
distances to determine whether the line strength can be used to estimate distances to faraway objects. They 
found several cases of significant column density difference in the interstellar component of stars 
in the same cluster or association, but were unable to determine whether this was caused by a local 
contribution to the derived profile, or confusion caused by background or foreground stars being mis-classified 
as cluster objects. Similar results were obtained by Smoker et al. (2011) who observed three open 
clusters with UVES and found, within the same cluster, variations in interstellar column density in Ca\,{\sc ii} and 
Na\,{\sc i} D of up to $\sim$0.5 and 1 dex, respectively. Authors such as Bowen et al. (1991), Kennedy et al. (1998), 
Meyer \& Lauroesch (1999), Andrews et al. (2001), Smoker et al. (2003, 2011), van Loon et al. (2009, 2013), Welsh et al. (2009) 
and Nasoudi-Shoar et al. (2010) attempted to eliminate the confusion issue by observing the centre of Globular clusters, QSOs or stars at high Galactic latitude 
in order to obtain the reduced Ca\,{\sc ii} or Na\,{\sc i} equivalent width. The current paper builds on some previous work by observing a total 
of 609 stars within 7 open clusters, of which 403 objects lie in the Large and Small Magellanic Clouds. By comparison with many previous studies, 
smaller scales ($\sim$12 arcsec to 27 arcmin corresponding to $\sim$0.05 to 15 pc) are studied, corresponding to 
the fieldsize of FLAMES\footnote{FLAMES (Pasquini et al. 2002) is a multi-object, intermediate and high
resolution spectrograph, mounted at the VLT/Unit Telescope 2 (Kueyen)
at Cerro Paranal, Chile, operated by ESO.} which is $\sim$30 arcmin, but still at a reasonable spectral resolution ($\sim$16 km\,s$^{-1}$). 

Additionally to the small-scale FLAMES observations, the large-scale structure of the Milky Way is studied using 
FEROS\footnote{FEROS (Kaufer et al. 1999) is a high-resolution echelle spectrograph,
mounted at the 2.2\,m Telescope at La Silla, Chile, operated by ESO.} and UVES\footnote{UVES (Dekker et al. 2000; 
Smoker et al. 2009) is a a high-resolution echelle spectrograph, mounted at the 8.2-m Unit Telescope 2 at Paranal, Chile}
via the observation in Ca\,{\sc ii} K and Na\,{\sc i} D of 165 stars in the Magellanic Clouds and 29 within the Galaxy at a 
resolution of $\sim$4--8 km\,s$^{-1}$. Our aim is to determine if current models of the ISM match with the observations. 

Section \ref{s_sample} describes the sample of open clusters and field stars, plus the data reduction and analysis performed to estimate 
the equivalent width and column density in the Galactic component in Ca\,{\sc ii} and/or Na\,{\sc i} D of the 
sightlines studied. In Sect. \ref{s_results} we give the main results, including figures showing the interstellar profiles of the 
sightlines studied as well as tables of the component fit parameters. Section \ref{s_discussion} contains the discussion, 
in particular focusing on the large- and small-scale variation observed in the Ca\,{\sc ii} K and Na\,{\sc i} D profiles towards the target stars and 
how variation in the former impacts on the use of the former species as a distance indicator. Using the FEROS and UVES Galactic sightlines plus 
archive data we determine the parallax -- column density relationship for Ca\,{\sc ii} and other lines and compare with the distance 
derived using spectroscopic parallax. Finally, Sect. \ref{s_summary} contains the summary and suggestions for future work. In 
what follows we define Low Velocity gas as having absolute values of velocity less than $\sim$35 km\,s$^{-1}$, intermediate velocity 
gas with $\sim$35$<v<\sim$90 km\,s$^{-1}$, and high velocity gas with absolute velocities greater than $\sim$90 km\,s$^{-1}$.

\section{The sample, data reduction and analysis}
\label{s_sample}

The data on which the current paper is based were extracted from the ESO archive and are FLAMES-GIRAFFE observations 
towards three open clusters located in the Milky Way, and two in each of the Large and Small Magellanic Clouds, plus FEROS 
and UVES observations towards stars located in the Magellanic system and Milky Way. 

\subsection{FLAMES-GIRAFFE archive sample towards open clusters in the MW and Magellanic System}
\label{s_sample_FLAMES}

The FLAMES data for the seven open clusters were taken from the ESO archive. Data reduction was performed using the 
ESO FLAMES pipeline within gasgano (Izzo et al. 2004) using calibrations taken on the day following the observation. For some observations 
the simultaneous calibration fibre was used. The HR2 grating was 
employed and measurements from a few spectra with the simultaneous calibration fibre enabled shows that the wavelength 
range was from $\sim$3850 to 4045\AA \, with a spectral resolution around Ca\,{\sc ii} K of $R$=$\lambda/\Delta\lambda\sim$18,500 
or $\sim$16 km\,s$^{-1}$. The re-binned data from the pipeline were imported into {\sc iraf}{\it \footnote{
{\sc iraf} is distributed by the National Optical Astronomy Observatories, U.S.A.}} where they were co-added using 
median addition within {\sc scombine} and then read into {\sc dipso} (Howarth et al. 1996). Subsequently, the spectra were 
normalised by fitting the continuum in regions bereft of interstellar features, and the signal-to-noise (S/N) ratio measured. 
At this stage a small number of late-type stars were excluded from the sample as in these cases distinguishing stellar from 
interstellar lines was problematical. For the early-type stars the stellar features were normally broad (c.f. Mooney et al. 2002, 2004) 
and hence easily removed in the normalisation process. Table \ref{t_sample} shows basic data on the open clusters for which FLAMES data were analysed. 
Columns 1--5 give the cluster name, alternative name, location (Galactic or Magellanic), Galactic coordinates of the plate centre and  
distance in kpc. Columns 6--8  give the total exposure time in seconds, median S/N ratio per pixel around Ca\,{\sc ii} K 
and the number of stars used in the analysis. This number only includes objects where the 
Galactic interstellar component was useful and varies from 43 usable spectra for NGC\,6611 to 110 for NGC\,330, or a total 
of 609 sightlines with Ca\,{\sc ii} interstellar measurements. Finally, columns 9--10 give the range of Galactic scales probed 
with the current measurements, being the minimum or maximum star-star separation at the distance of the cluster. For the 
Magellanic objects we assumed the scale height of the Galaxy in Ca\,{\sc ii} K is $\sim$800 pc (Smoker et al. 2003). 
Fig. \ref{f_FLAMES_Star_lb} shows the ($l,b$) coordinates of the stars used for each of the 
clusters. Tables \ref{t_FLAMES_NGC_330_sample} to \ref{t_FLAMES_NGC_6611_sample} (available online) show the coordinates and magnitudes of 
the individual stars on which the FLAMES-MEDUSA fibres were placed.

\begin{table*}
\caption{Open cluster basic data observed with FLAMES sorted in terms of increasing NGC number. Distances to Milky Way clusters 
are from the WEBDA database with the distances to the LMC and SMC being taken from Keller \& Wood (2006). The scales 
probed column corresponds to the minimum and maximum transverse star-star separation at the distance of the cluster. 
For the Galactic scales probed by the Magellanic objects we have assumed that the scaleheight of the Galaxy is 
$\sim$800 pc (Smoker et al. 2003). Note that the Galactic scales 
are upper limits, as the ISM absorption can arise anywhere in the 
line-of-sight between the Earth and the cluster in question.
}                               
\label{t_sample}                                                 
\centering                                                       
\begin{tabular}{l r r r r r r r r r}                           
\hline                                                         
Cluster    &  Alternative     &   Location   & ($l,b$)          &   Dist      &  Exp time &  Median S/N          & Stars  & Scales             &  Galactic scales \\
           &  name            &              &  (deg.)          &   (kpc)     &  (s)      &  at Ca\,{\sc ii} K   &  used  & probed (arcmin)    &  probed (pc)     \\
\hline                                                  
NGC\,330   &  Kron 35         &    SMC       & 302.42, --44.66  &   61        &  13650    &       30             &  111   & 0.2 -- 27.4        & 0.07 --  9.1     \\
NGC\,346   &  Kron 39         &    SMC       & 302.14, --44.94  &   61        &   6825    &       60             &  110   & 0.3 -- 20.7        & 0.12 --  8.9     \\
NGC\,1761  &  LH\,09          &    LMC       & 277.23, --36.07  &   51        &  13650    &      135             &  111   & 0.2 -- 22.3        & 0.09 --  8.8     \\
NGC\,2004  &  KMHK 991        &    LMC       & 277.45, --32.63  &   51        &  13650    &       95             &  111   & 0.2 -- 20.0        & 0.05 --  4.6     \\
NGC\,3293  &  Collinder 224   &    Galaxy    & 285.85,   +0.07  &   2.327     &    795    &       70             &   90   & 0.2 -- 22.3        & 0.14 -- 15.1     \\
NGC\,4755  &  The Jewel Box   &    Galaxy    & 303.21,   +2.50  &   1.976     &    795    &       80             &   81   & 0.2 -- 21.8        & 0.11 -- 12.5     \\
NGC\,6611  &  M\,16           &    Galaxy    &  16.98,   +0.80  &   1.749     &    750    &       50             &   43   & 0.2 -- 23.5        & 0.10 -- 12.0     \\
\hline                                 
\end{tabular}
\end{table*}

\subsection{FEROS and UVES archive data towards the Magellanic system}
\label{ferosplusuves}

FEROS and UVES spectra of stars within the Large and Small Magellanic Clouds from several observing runs were also 
extracted from the ESO archive. 

They are typically observations towards bright early-type O- and B- type stars. The Na\,{\sc i} D observations 
are sometimes affected by Na in emission, although this always appears slightly offset from the Galactic 
absorption and was not fitted when performing profile fits. 
Telluric correction around the Na\,{\sc i} D lines was performed by removing a scaled model of the sky smoothed to the spectral 
resolution of the observations using {\sc skycalc}\footnote{{\sc skycalc} is available at https://www.eso.org/observing/etc/skycalc}.

The aim of extracting these spectra was to determine the variation in the Ca\,{\sc ii} and Na\,{\sc i} D {\em Galactic} column density over large scales. 
Authors such as Megier et al. (2009) found large variations in the Ca\,{\sc ii} K line strength in Galactic open clusters, 
although could not rule out non-cluster contamination by background and foreground stars. The current observations of Magellanic 
targets eliminate the distance uncertainty and have median S/N ratios of $\sim$35 and $\sim$80 per pixel in Ca\,{\sc ii} and Na\,{\sc i}, respectively. 
Table \ref{t_FEROS_uves_Galactic_Magellanic} (available online) lists details of the 154 Magellanic Cloud targets observed. They are plotted in Fig. \ref{f_feros_Magellanic} 
and probe structures of size $\approx$5 degrees, and hence act as a useful comparison to FLAMES observations of LMC and SMC targets in Sect. \ref{s_sample_FLAMES} 
that probe scales of less than $\sim$30 arcminutes.

\begin{figure}
\setcounter{figure}{1}
\epsfig{file=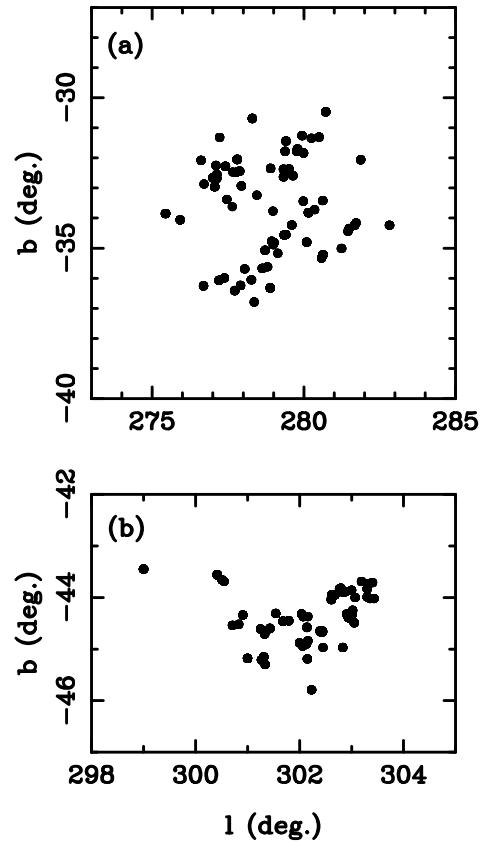}
\caption[]{Positions on the sky of stars taken from the FEROS and UVES archives used to investigate the large-scale structure 
variation of interstellar Galactic Ca\,{\sc ii} K and Na\,{\sc i} D. (a) LMC sample (b) SMC sample.}
\label{f_feros_Magellanic}           
\end{figure}

\subsection{FEROS archive data towards Galactic early-type stars}

Data towards Milky Way early-type stars from two observing runs were extracted from the FEROS archive. 
They are typically observations towards bright early-type O- and B- type stars. 
Authors such as Struve (1928), Megier et al. (2005, 2009)   
and references therein) have postulated the use of the Ca\,{\sc ii} K line strength as a distance indicator for objects close to the Galactic plane. 
Table \ref{t_FEROS_uves_Galactic_Magellanic} (available online) lists details of the Milky Way targets observed, with Fig. \ref{f_feros_Galactic_Only} 
showing their sky positions. The median S/N ratio was $\sim$200 and $\sim$230 in Ca\,{\sc ii} and Na\,{\sc i} D, respectively. 
In Sect. \ref{s_parallax_coldens} these observations are used with archive data to investigate the parallax -- column density relationship 
for Ca\,{\sc ii} K. 

\setcounter{figure}{2}
\begin{figure}
\epsfig{file=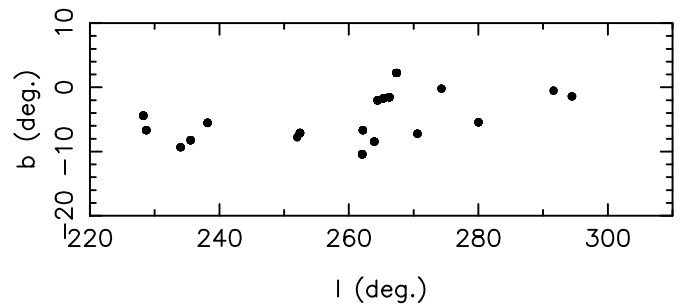}
\caption[]{Positions on the sky of data taken from the FEROS archive used to investigate the large-scale structure 
variation of Galactic Ca\,{\sc ii} K and Na\,{\sc i} D and the relationship between parallax and Ca\,{\sc ii} column density.}
\label{f_feros_Galactic_Only}           
\end{figure}

%
%

\subsection{Data analysis; component fitting}

For the Galactic absorption, interstellar components were fitted using both Gaussian fitting using the {\sc elf} 
routine within the {\sc dipso} software package, and also full profile Voigt-profile fitting using {\sc vapid} (Howarth et al. 2002). 

Simultaneous fitting was used for the Ca\,{\sc ii} H and K line profiles within the whole sample and for 
Na\,{\sc i} D$_1$ and D$_2$ in the FEROS and UVES datasets. The Ca\,{\sc ii} H line lies in the wing of H$\epsilon$, and was hence normalised to provide a 
profile which could be fitted simultaneously with Ca\,{\sc ii} K. Examples of Ca\,{\sc ii} H 
and K spectra are shown in online Fig. \ref{f_FLAMES_CaH_CaK_Examples}.

{\sc ELF} gives velocity centroids, full width half maxima and equivalent widths of the interstellar components. In addition, {\sc vapid} 
yields estimates of the $b$-values and column densities of the profiles. 
The wavelength for Ca\,{\sc ii} of 3933.661\AA \, and f-value of 0.627  were taken from Morton et al. (2003, 2004); conversion from 
Topocentric to the     Local Standard of Rest (LSR) reference frame was performed using {\sc rv} (Wallace \& Clayton 1996). We note that due to the 
spectral resolution of the FLAMES-GIRAFFE data ($\sim$16 km\,s$^{-1}$) it is likely that the interstellar profiles observed are 
in fact a superposition of many different components. For example, observations of Galactic gas in Ca\,{\sc ii} K by Welty 
et al. (1996) show that the vast majority of components in their sample have $b$-values of between 0.5--3.0 km\,s$^{-1}$, 
components that would be unresolved in the current dataset. In total Galactic interstellar profiles towards 609 stars were fitted. Smoker et al. (2015) separately describe the 
analysis of intermediate and high velocity clouds observed in the spectra towards the Magellanic Cloud targets. 

Errors in the interstellar components were estimated using procedures outlined in Hunter et al. (2006). Briefly, 
these involve changing the column density and $b$-value of each component in turn until the residual in the model-data exceeds 
1$\sigma$ in 3 adjacent velocity bins. For the cases where no residual was above the limit even when the change 
equaled the measurement value, the error was set to the value of the measurement. This most frequently happened in 
the components with small $b$-values. 

Two of the FEROS sample stars (HD 53244 and HD 76728) have very strong stellar lines around Ca\,{\sc ii} and Na\,{\sc i} hence 
profile fitting was not performed for these objects.

\section{Results}
\label{s_results}

In this section we present the reduced spectra and model fits for the FLAMES-GIRAFFE and FEROS/UVES sample. 

\subsection{FLAMES-GIRAFFE spectra}

Figures \ref{f_NGC_330_CaK_FLAMES_Spectra} to \ref{f_NGC_6611_CaK_FLAMES_Spectra} (available online) show the 
Ca\,{\sc ii} K spectra towards each of the stars in the sample as well as the model fit obtained using Gaussian (ELF) fitting and the 
(data-model) residual fit. In order to assess the variations in the profiles, Figs. \ref{f_MaxVariation_EW_MW_NGC_330} to 
\ref{f_MaxVariation_EW_MW_NGC_6611} (also available online) show the 16 star-star pairs in each cluster with the largest 
differences in equivalent width, with no star being plotted more than once.
Tables \ref{t_NGC_330_cmpt_fits} to \ref{t_NGC_6611_cmpt_fits} (available on-line) show the corresponding 
Voigt profile fit results for each of the seven clusters studied. 

\subsection{FEROS and UVES spectra}

Figure \ref{f_FEROS_MC_Gal_CaK_NaD_HI} (available online) shows the FEROS and UVES Ca\,{\sc ii} K, Na\,{\sc i} D and 
corresponding GASS and LABS Survey 21-cm H\,{\sc i} (Kalberla et al. 2005; McClure-Griffiths et al. 2009) 
spectra towards the 165 Magellanic Cloud and 29 Milky Way stars. The latter
data have velocity resolution of $\sim$1 km\,s$^{-1}$, brightness temperature sensitivity of 0.06--0.07 K and 
spatial resolution of $\sim$0.5$^{\circ}$ (LAB) and 16 arcmin (GASS). Tables A16 and A17 
show the corresponding profile fits and total column densities derived from the optical data, plus the total derived H\,{\sc i} Galactic column density  
derived from the equation; $N_{\rm HI}$=1.823$\times$10$^{18}\times \int T_{\rm B}$ $dv$, where $T_{\rm B}$ 
is the detected brightness temperature and $dv$ is in km\,s$^{-1}$. 

Figures \ref{f_FEROS_CaK_Galactic_MaxNDiff} and \ref{f_FEROS_NaD_Galactic_MaxNDiff} show the 16 Magellanic
sightlines pairs in Ca\,{\sc ii} K and Na\,{\sc i} D for which there is the greatest difference 
in column density. Each sightline is only plotted once. 

%
%

\section{Discussion}
\label{s_discussion}

In this section we discuss the composite Ca\,{\sc ii} K spectra and their comparison with single-dish H\,{\sc i} 21-cm observations, 
the reduced equivalent widths and column densities for the local Ca\,{\sc ii} gas as a function of sky position, the variation 
in the equivalent width with sky position and the velocity structure, the Ca/H\,{\sc i} and Na/H\,{\sc i} ratios as a function 
of $N$(H\,{\sc i}), the Ca\,{\sc ii}/Na\,{\sc i} ratio and finally 
the parallax -- Ca\,{\sc ii} column density relationship. 

\subsection{Composite GIRAFFE Ca\,{\sc ii} K spectra and comparison with 21-cm H\,{\sc i} observations}
\label{s_composite}

Figure \ref{f_Cluster_composite_spectra_CaK_HI} shows the {\em composite} Ca\,{\sc ii} K spectra towards each of the seven clusters, formed by median-combining the 
individual normalised spectra, weighting by the square of the S/N ratio, and boxcar smoothing using a box of 3 pixels. We note that the FWHM of the arc lines 
is 4 pixels, so no degradation in resolution occurs due to the smoothing. The composite spectra have S/N ratios ranging from $\sim$500-1200, and display between 1-2 main 
components with velocities from  --35 to +35 km\,s$^{-1}$. Also shown in Fig. \ref{f_Cluster_composite_spectra_CaK_HI}  are 21-cm H\,{\sc i} data taken from 
the LABS and GASS surveys. Tables \ref{t_CaK_composite_fits} and \ref{t_HI_composite_fits} show the results of component fitting 
to the composite Ca\,{\sc ii} K and single-dish H\,{\sc i} spectra, with values of the abundance $A$=log($N$(Ca)/$N$(H\,{\sc i})) given. 
Although the ATCA-Parkes H\,{\sc i} survey (Kim et al. 2003) of the Magellanic Clouds covers the LMC, 
no velocity information is available for the Galactic component. 

For the three Galactic clusters the H\,{\sc i} in emission has significantly more velocity structure than the Ca\,{\sc ii} in absorption, which is 
simply a reflection of the difference in path lengths studied, with the clusters being at distances of $\sim$2-kpc compared with the extent of the 
Milky Way disc in H\,{\sc i} that extends beyond a radius of 40-kpc (e.g. Kalberla et al. 2007 and references therein). For three of the four 
Magellanic clusters, the main low-velocity peaks observed in the H\,{\sc i} spectra are also visible in the Ca\,{\sc ii} data. The 
exception is NGC\,2004 for which there are two bright H\,{\sc i} components separated by $\sim$8 km\,s$^{-1}$ plus an IV component at +40 km\,s$^{-1}$ 
but for which only one Ca\,{\sc ii} component is detected at a spectral resolution of $\sim$16 km\,s$^{-1}$. This Ca\,{\sc ii} feature has a 
FWHM (corrected for instrumental broadening) of $\sim$27 km\,s$^{-1}$,  indicative of two or more components (otherwise the kinetic temperature 
of the gas would be extremely high). 

We comment finally on the NGC\,1761 interstellar spectra in H\,{\sc i} and Ca\,{\sc ii} K. In H\,{\sc i} there is a weak emission 
feature at $\sim$--20 km\,s$^{-1}$ and another much stronger one at $\sim$3 km\,s$^{-1}$. The component at --20 km\,s$^{-1}$ has a 
[Ca \,{\sc ii}/H\,{\sc i}] equivalent width/column density ratio approximately 4 times larger than that at +3 km\,s$^{-1}$. This 
increasing ratio with velocity is frequently seen and is generally thought to be caused by calcium being 
liberated from dust into the gas phase in intermediate- and high-velocity gas (e.g. Wakker \& Matthis 2000). 

\begin{table}
\setcounter{table}{1}
\caption{Ca\,{\sc ii} ELF fit results for the composite FLAMES-GIRAFFE spectra. Velocities are in the LSR. FWHM velocities are observed and 
do not take into account instrumental broadening of $\sim$16 km\,s$^{-1}$. Equivalent widths of the Ca\,{\sc ii} K lines are 
in m\AA. See Sect. \ref{s_composite} for details.
}                               
\label{t_CaK_composite_fits}                                     
\centering                                                       
\begin{tabular}{l r r r}                                      
\hline                                                           
Cluster    &        $v$        &   FWHM         &  $EW$         \\
           &   (km\,s$^{-1}$)  & (km\,s$^{-1}$) &  (m\AA)       \\
\hline                                          
NGC\,330   &        -0.3       &    24.9        &        93.6   \\
   "       &        63.0       &    35.8        &        36.4   \\
   "       &       118.8       &    23.1        &        97.9   \\
   "       &       133.4       &    57.4        &       126.1   \\
   "       &       202.3       &    14.9        &         2.2   \\
   "       &       352.9       &    20.1        &         5.0   \\
NGC\,346   &         0.0       &    24.6        &        95.8   \\
   "       &        84.7       &    46.8        &        58.2   \\
   "       &       117.3       &    26.5        &        75.4   \\
   "       &       140.5       &    24.0        &        90.8   \\
   "       &       160.2       &    34.5        &        63.2   \\
   "       &       202.2       &    18.5        &         3.9   \\
NGC\,1761  &       -41.9       &    12.5        &         2.2   \\
   "       &       -19.1       &    20.1        &        45.1   \\
   "       &         1.6       &    18.1        &        56.9   \\
   "       &        46.6       &    16.3        &         3.3   \\
   "       &        76.1       &    28.6        &        17.0   \\
   "       &       142.1       &    26.7        &        12.8   \\
   "       &       205.0       &    62.1        &        12.1   \\
   "       &       252.8       &    47.6        &        66.7   \\
   "       &       263.4       &    24.4        &        69.9   \\
 NGC\,2004 &       -18.8       &    31.4        &         1.4   \\
    "      &         0.5       &    19.9        &        72.0   \\
    "      &        46.5       &    22.6        &        23.0   \\
    "      &       102.1       &    25.5        &        19.7   \\
    "      &       239.8       &    33.8        &        46.8   \\
    "      &       283.2       &    38.7        &        24.1   \\
    "      &       334.4       &    35.0        &         4.1   \\
 NGC\,3293 &       -86.8       &    77.4        &         9.2   \\
    "      &       -48.6       &    19.7        &         5.5   \\
    "      &        -9.6       &    30.9        &       276.1   \\
    "      &         9.9       &    46.6        &        18.5   \\
 NGC\,4755 &       -44.3       &    23.6        &        39.9   \\
    "      &       -21.3       &    26.0        &       215.9   \\
    "      &        -1.6       &    18.5        &        96.8   \\
 NGC\,6611 &       -48.5       &    74.2        &        12.6   \\
    "      &       -28.6       &    14.2        &         3.9   \\
    "      &         2.4       &    14.1        &        24.9   \\
    "      &        15.8       &    41.2        &       478.3   \\
    "      &        52.5       &    36.6        &        31.5   \\
\hline
\end{tabular}
\end{table}

\begin{table}
\setcounter{table}{2}
\caption{H\,{\sc i} fit results for the LABS spectrum at the coordinate of the centre of the FLAMES-GIRAFFE pointing for 
the Magellanic objects. 
See Sect. \ref{s_composite} for details.
}                               
\label{t_HI_composite_fits}                                      
\centering                                                       
\begin{tabular}{l r  r r r}                             
\hline                                                           
Cluster    &        $v$        &   FWHM         & $T_{\rm peak}$   &  $N_{\rm HI}$ \\
           &   (km\,s$^{-1}$)  & (km\,s$^{-1}$)  &  (K)             &  (cm$^{-2}$)  \\
\hline                                                  
  NGC\,330 &        -0.5       &     5.4        &        10.5       &        20.0   \\
     "     &         1.7       &    16.7        &         6.1       &        20.3   \\
     "     &       114.8       &    30.3        &        22.3       &        21.1   \\
     "     &       123.3       &    19.6        &        34.8       &        21.1   \\
     "     &       159.5       &    30.6        &        49.2       &        21.5   \\
  NGC\,346 &        -0.4       &     6.1        &         7.5       &        19.9   \\
     "     &         0.0       &     2.5        &         5.2       &        19.4   \\
     "     &         0.3       &    20.0        &         4.2       &        20.2   \\
     "     &        99.1       &    15.3        &         1.4       &        19.6   \\
     "     &       122.9       &    20.8        &        18.8       &        20.9   \\
     "     &       158.5       &    23.4        &        47.1       &        21.3   \\
     "     &       173.3       &    24.1        &        24.1       &        21.1   \\
  NGC\,1761&       -19.9       &    11.3        &         1.8       &        19.6   \\
     "     &         3.0       &    18.4        &         7.5       &        20.4   \\
     "     &         5.0       &     6.2        &         3.7       &        19.7   \\
     "     &       256.8       &    12.0        &         1.5       &        19.6   \\
     "     &       278.7       &    44.5        &        10.1       &        20.9   \\
     "     &       280.1       &    21.3        &        36.5       &        21.2   \\
  NGC\,2004&        -3.1       &    12.9        &        11.3       &        20.5   \\
     "     &         5.9       &     6.1        &         7.0       &        19.9   \\
     "     &         7.4       &    14.8        &         3.4       &        20.0   \\
     "     &       269.2       &    43.5        &         3.5       &        20.5   \\
     "     &       297.7       &    22.6        &         8.3       &        20.6   \\
     "     &       287.2       &    16.5        &         6.8       &        20.3   \\
     "     &       331.6       &    23.4        &         0.7       &        19.5   \\
\hline
\end{tabular}
\end{table}

   \begin{figure*}
   \setcounter{figure}{3}
   \centering
   \includegraphics{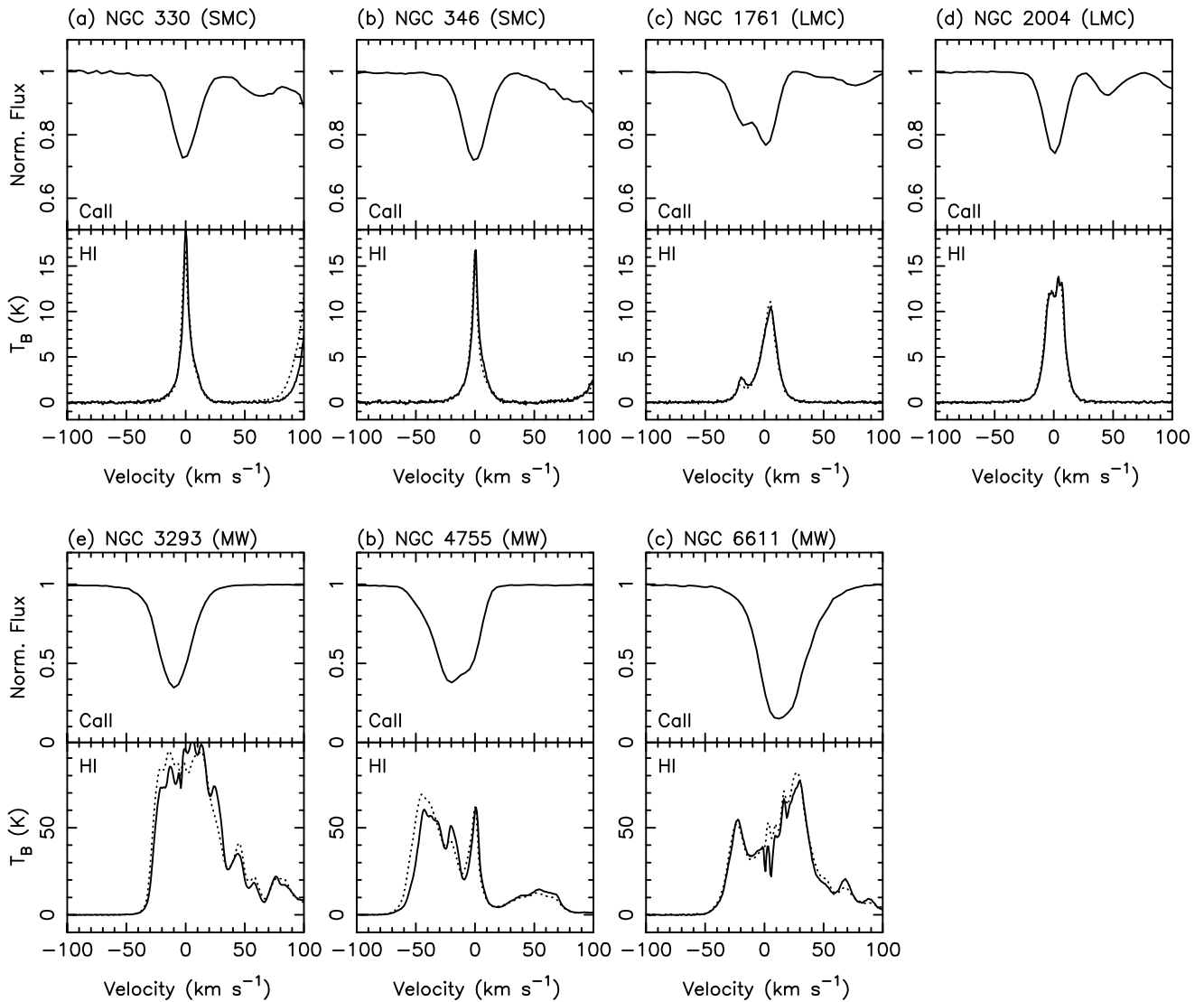}
   \caption{Top panels: Composite Ca\,{\sc ii} K spectra towards each of the 7 clusters, formed by median combining the individual fibres after 
            continuum normalisation. Bottom panels: 21-cm H\,{\sc i} data taken from the GASS (full lines) and LABS survey (dotted lines) towards the FLAMES plate centres. 
            (a) NGC\,330 (S/N$\sim$500), (b) NGC\,346 (S/N$\sim$900), (c) NGC\,1761 (S/N$\sim$1200), (d) NGC\,2004 (S/N$\sim$1000), 
            (e) NGC\,3293 (S/N$\sim$800), 
            (f) NGC\,4755 (S/N$\sim$800), (g) NGC\,6611 (S/N$\sim$500). Absorption-line features at $\sim$50 km\,s$^{-1}$ are caused by IV gas.}
   \label{f_Cluster_composite_spectra_CaK_HI}
   \end{figure*}


\subsection{Reduced equivalent widths and column densities for Ca\,{\sc ii} K and Na\,{\sc i} D and comparison with previous work}
\label{s_rews}

Figure \ref{f_Variation_N_vs_position_P1} (available online) 
shows the point-to-point variation in total column density and 
percentage difference in equivalent width as a function of transverse separation for each of the 7 clusters observed with FLAMES. The 
column densities and equivalent widths were integrated between the velocity limits shown on the figures in order to 
exclude intermediate-, high- and Magellanic Cloud velocity components. We again note that due to the relatively low spectral resolution 
of the FLAMES-GIRAFFE dataset, unresolved components are likely to be present that make the column densities very 
uncertain. 

Bowen et al. (1991) and Smoker et al. (2003) find reduced equivalent width (REW) values in the Ca\,{\sc ii} K line for objects at 
infinity of $\sim$110 m\AA \, (with 95 percent of lines lying between 60-310 m\AA) and 113 m\AA, respectively, where the REW 
is defined as EW$\times$sin($b$). In the current FLAMES-GIRAFFE dataset for the 4 Magellanic clusters in Ca\,{\sc ii} K we find REW values for Galactic 
gas ranging from $\sim$35--125 m\AA\ (see Table \ref{t_REW_CaK_NaD}), with median values of $\sim$40--70 m\AA \, on scales of $\sim$0.05--6 pc. 
For the individual stars observed by FEROS and UVES, the corresponding range is $\sim$30--125 m\AA \, in Ca\,{\sc ii} K and $\sim$50--155 m\AA \, 
in Na\,{\sc i} D, with median values of 45 and 100 m\AA, respectively. Figure \ref{f_Histogram_REW_RlogN_CaK_FLAMES} 
shows histogrammes of the REW for low-velocity gas observed in absorption towards four Magellanic clusters, with 
Fig. \ref{f_Histogram_REW_CaK_Gal_FLAMES} showing the corresponding histogrammes for the Galactic clusters. For the MC sightlines the 
reduced equivalent widths are approximately half the values of those observed in previous work (Bowen et al. 1991), and again indicate large-scale variations 
in the EW of optical absorption lines.  

For the FLAMES-GIRAFFE sightlines, the maximum variation in REW in low-velocity gas over the $\sim$5 pc field of view is a 
factor 3.0 for NGC\,330 (which has lower S/N ratio than the other sightlines), 1.8 for NGC\,346, 1.8 for 
NGC\,1761 and 1.6 for NGC\,2004. These variations are somewhat smaller than observed in the intermediate velocity and high velocity gas towards 
the same sightlines, where the Ca\,{\sc ii} K REW for example towards NGC\,2004 
varies by factors exceeding 10 (Smoker et al. 2015). Of course, the IV gas is likely to be at larger distances than the LV gas, hence the transverse 
scales sampled are bigger. Previous studies of small-scale ($\sim$0.03 pc) structure using binaries or the cores of Globular 
Clusters (e.g. Meyer \& Blades 1996, Lauroesch \& Meyer 1999, Lauroesch 2007 and references therein) have found 
strong variations in Na\,{\sc i} D profiles on small scales, but much smaller changes in Ca\,{\sc ii} K equivalent widths or 
column densities. The current observations confirm that such equivalent width variations in Ca\,{\sc ii} also exist on  
scales of $\sim$0.05--6 pc, with variation of $\sim$0.3--0.5 dex in the optically thin approximation. 

Figure \ref{f_REW_RLogN_Hist_FEROS_CaK_NaD} shows histograms of the reduced equivalent width and reduced 
column density for the two FEROS/UVES-observed species. 
For the LMC-only 
sightlines, 68 percent of the Ca\,{\sc ii} reduced column densities lie within $\pm$0.16 dex of log[$N$(Ca\,{\sc ii} cm$^{-2}$)]=11.85. 
For the Na\,{\sc i} data, 68 percent of the reduced column densities lie within $\pm$0.32 dex of log[$N$(Na\,{\sc i} cm$^{-2}$)]=11.93, 
reflecting the generally higher clumpiness of this neutral species compared with Ca\,{\sc ii}. 
Finally, Fig. \ref{f_REW_BigSmall_FEROS_CaK_NaD} shows the sightlines that display the biggest and smallest reduced 
equivalent width values for the low velocity gas in Ca\,{\sc ii} K and Na\,{\sc i} D for the Magellanic Cloud targets. 

For over fifty years astronomers have thought that hierarchical structures and turbulence exist in the ISM (von Weizsacker 1951; 
von Hoerner 1951; see reviews by Elmegreen \& Scalo 2004, Dickey 2007, Hennebelle \& Falgarone 2012 and Falceta-Goncalves et al. 2014). 
The power spectrum of the ISM has often been used to provide coarse-scale information on  
structures present in the ISM and indicate how much material is present at each scale. However, it provides
little information about the 
shape of the structures themselves, i.e. very different structures can produce similar power spectra (Chappell \& Scalo 2001). In any 
case, with the incompletely-sampled FLAMES and FEROS data we cannot obtain a reliable power spectrum of the column density variations. Hence we 
restrict ourselves to a comparison between our work and the similar observational and theoretical study of Van Loon et al. (2009). 
Their observations towards $\omega$ Cen found that the real fluctuations in the column density maps over scales 
of half a degree were 7 per cent in Ca\,{\sc ii} (1 standard deviation). The fluctuations detected 
in Ca\,{\sc ii} K are shown in Table \ref{t_REW_CaK_NaD} for our seven clusters. In the case of 
the Magellanic clusters,
the 1$\sigma$ variation in equivalent width for the FLAMES-observed field-of-view ranges from 9--15 per cent (for NGC\,330 and
NGC\,346, respectively) for Ca\,{\sc ii} in the gas phase with velocities between -35 and +35 km\,s$^{-1}$. For the three Galactic 
clusters the variation is 63 percent (NGC\,3293), 10 per cent (NGC\,4755) and 15 per cent for NGC\,6611. These are upper 
limits, not taking into account the errors on the measurements. For the FEROS/UVES spectra which span tens of degrees on the 
plane of the sky, the variation in column density is unsurprisingly much larger, being $\sim$51 per cent in Ca\,{\sc ii}. 

Van Loon et al. (2009) presented a simple model of the ISM as a collection of spherical cloudlets with filling factor 0.3 and sizes 
between 1 AU and 10 pc (their Appendix B). The model predicts observed fluctuations in the column density of the ISM on 
scales of 0.5 degrees of 0.1--0.2, 
consistent with our observed values. However, other physical 
forms of the ISM such as sheets or filaments may also be consistent with the observed variations (e.g. Heiles 1997, 
G\'{o}mez \& V\'{a}zquez-Semadeni 2014 and refs. therein). Indeed, Herschel observations of 
molecular clouds have detected 
a wealth of filaments towards the Gould Belt, with typical widths of around 0.1 pc (Andr\'e et al. 2010), although to our knowledge 
there are no existing optical absorption line data that show filaments of such size in the warm ISM.
\begin{table}
\setcounter{table}{3}
\begin{center}
\setcounter{table}{3}
\caption[]{Reduced equivalent width values for Galactic (low-velocity) gas using stellar probes. Minimum, maximum, 
median and 1-sigma variations values are shown for each dataset. The cluster objects are FLAMES-GIRAFFE observations and
the field objects FEROS and UVES observations.}
\label{t_REW_CaK_NaD}
\begin{tabular}{lrrrrr}
\hline
     Cluster      &     Species    &    REW$_{\rm Min}$   &   REW$_{\rm Max}$   &  REW$_{\rm Med}$ &  $\sigma$ \\
                  &                &      (m\AA)          &      (m\AA)         &       (m\AA)     &  (m\AA)   \\
\hline
  NGC\,330        &  Ca\,{\sc ii}  &         36.9         &      111.3          &       64.5       &   10.0    \\
  NGC\,346        &  Ca\,{\sc ii}  &         46.7         &       87.8          &       68.0       &    6.3    \\
  NGC\,1761       &  Ca\,{\sc ii}  &         46.1         &       81.6          &       60.2       &    7.6    \\
  NGC\,2004       &  Ca\,{\sc ii}  &         28.3         &       49.1          &       39.9       &    7.9    \\
  NGC\,3293       &  Ca\,{\sc ii}  &         0.03         &       1.54          &       0.41       &   0.26    \\
  NGC\,4755       &  Ca\,{\sc ii}  &         12.0         &       26.2          &       15.3       &    1.5    \\
  NGC\,6611       &  Ca\,{\sc ii}  &          4.8         &       11.8          &        7.8       &    1.2    \\
  Mag. field      &  Ca\,{\sc ii}  &         27.9         &      138.8          &       56.8       &   29.3    \\
\hline
\end{tabular}
\end{center}
\end{table}

\begin{figure*}
\setcounter{figure}{4}
\centering
\includegraphics{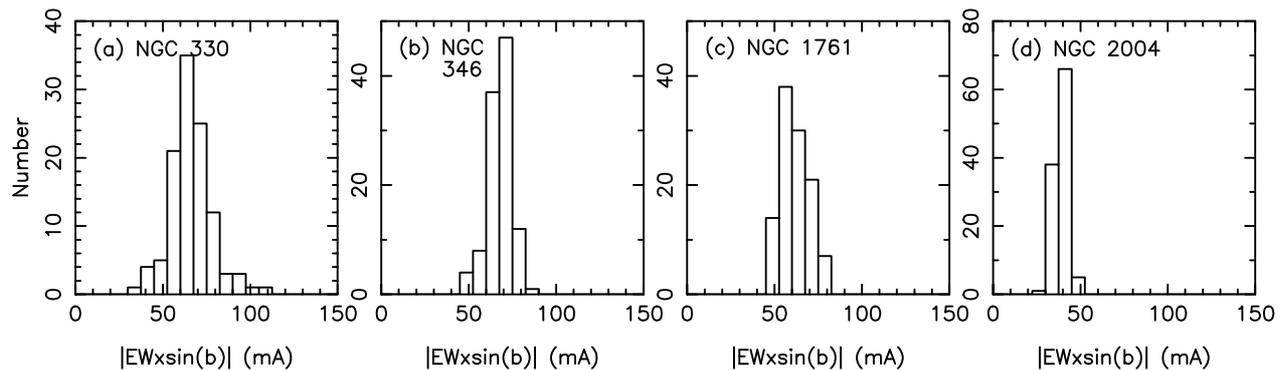}
\caption{Panels (a)-(d) Histogrammes of reduced equivalent width in the Ca\,{\sc ii} K interstellar line for Magellanic open 
clusters NGC\,330, NGC\,346, NGC\,1761 and NGC\,2004. The integration limits are the same as in Fig. \ref{f_Variation_N_vs_position_P1}. }
\label{f_Histogram_REW_RlogN_CaK_FLAMES}
\end{figure*}

\begin{figure*}
\setcounter{figure}{5}
\centering
\includegraphics{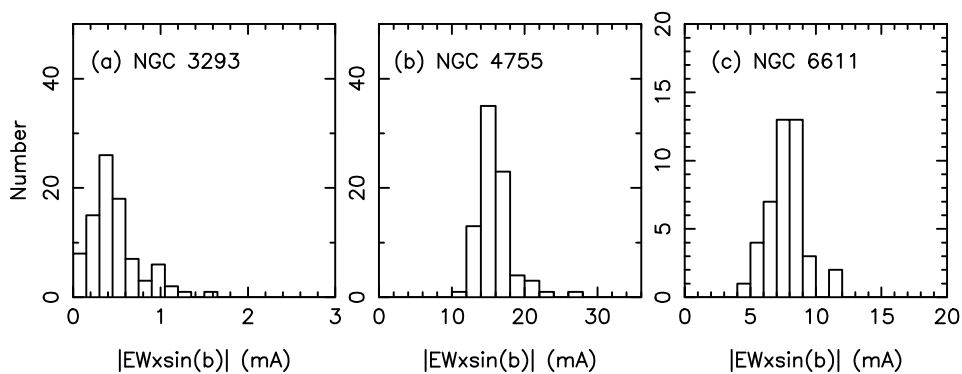}
\caption{Panels (a)-(c) Histogrammes of reduced equivalent width in the Ca\,{\sc ii} K interstellar line for Milky Way 
clusters NGC\,3293, NGC\, 4755, NGC\,6611. The integration limits are the same as in Fig. \ref{f_Variation_N_vs_position_P1}}
\label{f_Histogram_REW_CaK_Gal_FLAMES}
\end{figure*}

\begin{figure}
\setcounter{figure}{6}
\epsfig{file=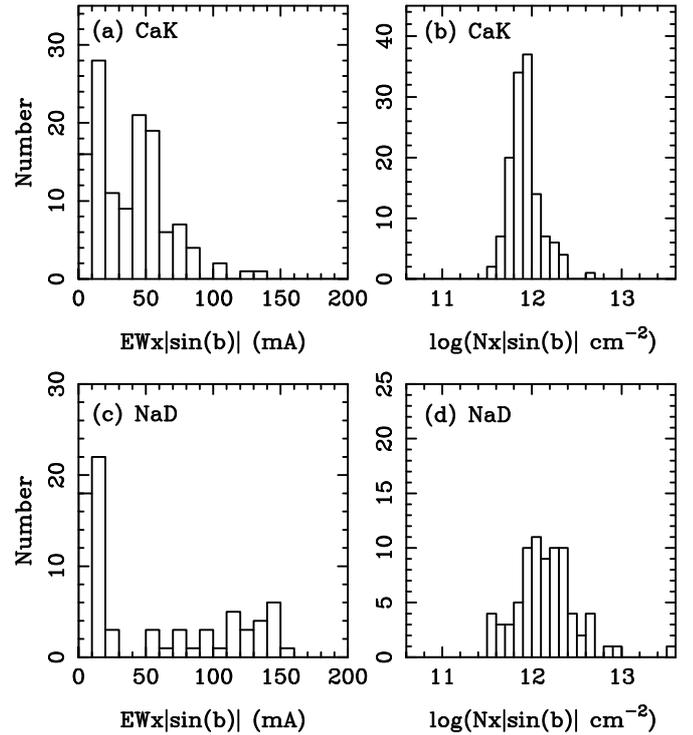}
\caption[]{Histogrammes showing the reduced equivalent width (REW) (panels (a) and (c)), and reduced column density  
(panels (b) and (d)) for Ca\,{\sc ii} K and Na\,{\sc i} D data from the Magellanic FEROS and UVES sample for gas with LSR velocities  
between --35 and +35 km\,s$^{-1}$. Note that a number of targets do not have REW measurements due to the presence of 
stellar lines. See Table 9 for details.} 
\label{f_REW_RLogN_Hist_FEROS_CaK_NaD}
\end{figure}

\begin{figure}
\setcounter{figure}{7}
\includegraphics[]{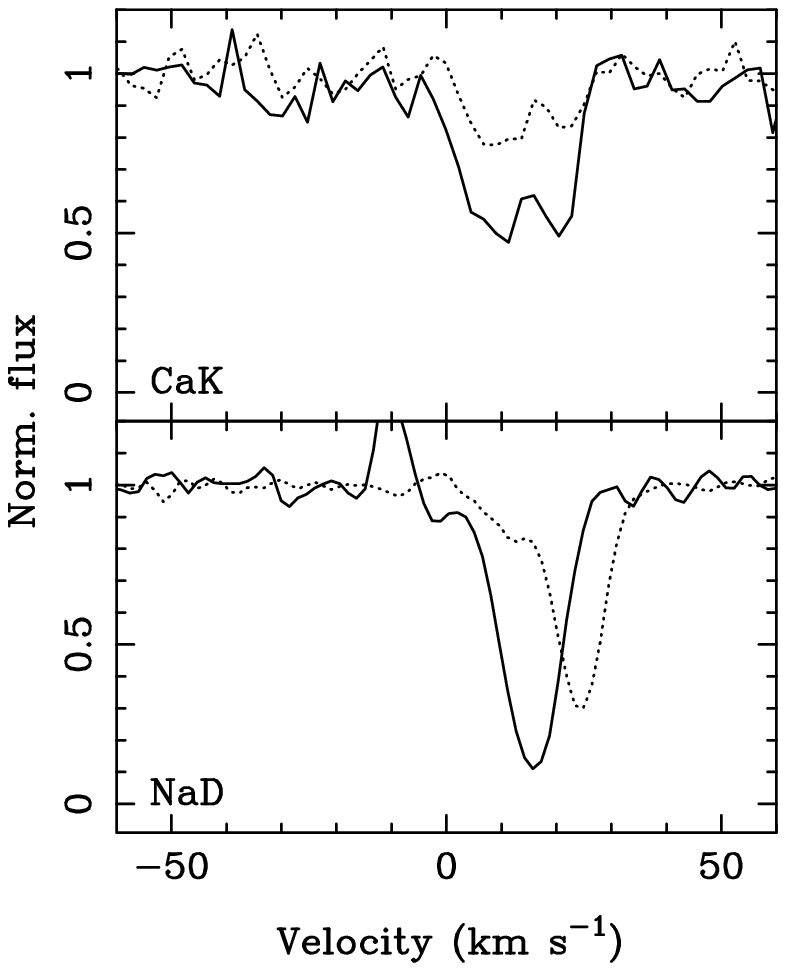}
\caption[]{FEROS and UVES spectra. Top panel: Ca\,{\sc ii} spectra of Magellanic targets for the sightlines that have largest
and smallest reduced equivalent widths between --35 and +35 km\,s$^{-1}$. One star is from the SMC and one from the LMC 
although we are probing Galactic gas. Bottom panel: Corresponding data
for the same two stars for Na\,{\sc i} D. Filled lines: LHA 115-S 23 (SMC). Dotted lines: SK -67 256 (LMC).} 
\label{f_REW_BigSmall_FEROS_CaK_NaD}
\end{figure}


\subsection{Variation in the Galactic velocity centroid and component structure} 
\label{s_velstructure}

Figures \ref{f_Velocity_EW_MW_NGC_330} to \ref{f_Velocity_EW_MW_NGC_6611} (available online) show the velocity centroid of the 
main Galactic Ca\,{\sc ii} K component for each of the clusters observed with FLAMES-GIRAFFE. There are hints of gradients in the velocity 
centroid for this low-velocity component in the GIRAFFE data only towards NGC\,1761 (north to south in Galactic coordinates with a magnitude 
of a few km\,s$^{-1}$) and towards NGC\,2004 (north-west to south-east of a few km\,s$^{-1}$  over a 0.5 degree field). These probe Galactic gas with maximum transverse 
scales of $\sim$5 pc. Figures \ref{f_FEROS_LMC_SMC_EW_Velocity_CaK} and \ref{f_FEROS_LMC_SMC_EW_Velocity_NaD} show the corresponding plots for the 
LV gas for the Magellanic stars observed with FEROS. 
The SMC clusters NGC\,330 and NGC\,346 
are generally well fitted by only one component at low velocities, although in particular for NGC\,330 there are sometimes indications 
of two-component structure that would need a better S/N ratio and/or spectral resolution to resolve. Considering the LMC clusters,
for NGC\,1761 there are generally two strong low-velocity components, and one for NGC\,2004 (although both frequently show 
intermediate and high-velocity gas, discussed in Smoker et al. 2015). For the Galactic clusters, NGC\,3293 can often be 
fit with a single component (e.g. Star 2372 in Fig. \ref{f_NGC_3293_CaK_FLAMES_Spectra}), although the residuals in other 
sightlines (e.g. Star 2341) imply another component may be needed, and in yet other sightlines (e.g. Star 2303) there is  
clearly more than one component. Nevertheless, the overall shape of the Ca\,{\sc ii} K profile is similar in all sightlines. More 
variation in profile shape is apparent towards Galactic cluster NGC\,4755, with all sightlines needing two or three components 
to be well-fitted. Finally, towards NGC\,6611 there is also a large variation in the two to four interstellar components present. 

%

\subsection{Ca\,{\sc ii}/H\,{\sc i} and Na\,{\sc i}/H\,{\sc i} ratios in the Galactic ISM from observations of stars in the Magellanic Clouds}
\label{s_element_ratios}

Values of the estimated Galactic Ca\,{\sc ii} and Na\,{\sc i} abundances $A$ were derived from log($N_{\rm opt}$)/log($N_{\rm HI})$ 
(where N$_{\rm opt}$ is the column density of either Ca\,{\sc ii} or Na\,{\sc i}). 
Figures \ref{f_FEROS_Ca_Na_vs_HI}(a) and (b) show the corresponding fits of $A$ against $N_{\rm HI}$ for Ca\,{\sc ii} 
and Na\,{\sc i} respectively, with the best-fitting lines from Wakker \& Mathis (2000)  
overlaid. We note that Wakker \& Mathis only plotted data up to log[$N$(H\,{\sc i} cm$^{-2}$)]$\sim$21.4, and that 
for the current dataset ionisation effects have not been taken into account. Neither has H$_{2}$ nor 
the large difference in resolution between the optical and H\,{\sc i} observations. For both Ca\,{\sc ii} and Na\,{\sc i}, 
at column densities smaller than log[$N$(H\,{\sc i})]=21.4, the values of $A$ lie within the 1$\sigma$ scatter 
of 0.42 and 0.52 dex, respectively, given in Wakker \& Mathis (2000). 
However, at higher H\,{\sc i} column densities the extrapolation of the best fit of Wakker \& Mathis (2000) lies $\sim$1 dex above 
the observed $A$ values for Ca\,{\sc ii}. Also, the scatter in $A$ for Na\,{\sc i} markedly increases at high values of H\,{\sc i}. This 
may be caused in part by saturation effects, and ideally the current sightlines should be re-observed in Na\,{\sc i} at 3303\AA\ to 
eliminate this possibility. 

\begin{figure}
\setcounter{figure}{8}
\includegraphics{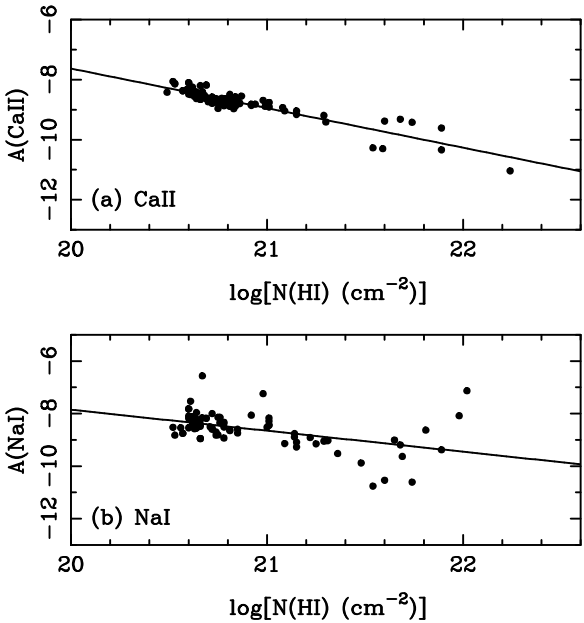}
\caption[]{$A$ vs $N$(H\,{\sc i}) for FEROS and UVES Magellanic sightlines for gas with LSR velocities between --35  
and +35 km\,s$^{-1}$. The full lines are from Wakker \& Mathis (2000) (a) Ca\,{\sc ii} K. (b) Na\,{\sc i} D.}
\label{f_FEROS_Ca_Na_vs_HI}
\end{figure}

\subsection{The Ca\,{\sc ii}/Na\,{\sc i} ratio in the FEROS/UVES Magellanic sightlines}

For LV gas the [Ca\,{\sc ii}/Na\,{\sc i}] ratio ranges from $\sim$--0.9 to +0.6 dex which is within the range of --0.1 to 100 
derived for example by Siluk \& Silk (1974) and Vallerga et al. (1993). The Ca\,{\sc ii}/Na\,{\sc i} column density ratio 
is a common diagnostic of the ISM (Hobbs 1975; Welty et al. 1999; van Loon et al. 2009; Welsh et al. 2009 amongst others), due to the fact 
that Ca shows a large range in depletion, depending on the temperature and presence of dust (e.g. Bertin et al. 1993).
Welty et al. (1999) note that if Ca\,{\sc ii} is the dominant species, then the ratio [Ca\,{\sc ii}/Na\,{\sc i}] 
depends primarily on the Ca depletion and the temperature. In warm gas (T$\sim$3000 K), Ca\,{\sc i} is enhanced and Ca\,{\sc iii} 
can also be a major contributor to the total Ca column density (Sembach et al. 2000). 
Figure \ref{f_FEROS_CaII_div_NaI_vs_CaII} shows this 
ratio plotted against Ca\,{\sc ii} K column density for the FEROS and UVES Magellanic sightlines only. A weak trend in increasing [Ca\,{\sc ii}/Na\,{\sc i}] 
with Ca\,{\sc ii} K column density is present, although this could be in part explained by saturation issues. A future paper will look at these 
and other data in more detail to investigate the [Ca\,{\sc ii}/Na\,{\sc i}] ratio as a function of velocity (the Routly-Spitzer effect; 
Routly \& Spitzer 1952, Vallerga et al. 1993). 

Figure \ref{f_FEROS_CaII_div_NaI_vs_HI} shows the FEROS and UVES-observed [Ca\,{\sc ii}/Na\,{\sc i}] ratio for low-velocity gas, derived using 
Magellanic objects, as a function of H\,{\sc i} column density for LV gas as obtained from the LABS survey. There is an anti-correlation 
between the two quantities, explained by the fact that the neutral species Na\,{\sc i} and H\,{\sc i} both probe cooler parts of the ISM than
Ca\,{\sc ii} which tends to be depleted onto dust as the H\,{\sc i} column density increases. 


\begin{figure}
\setcounter{figure}{9}
\includegraphics{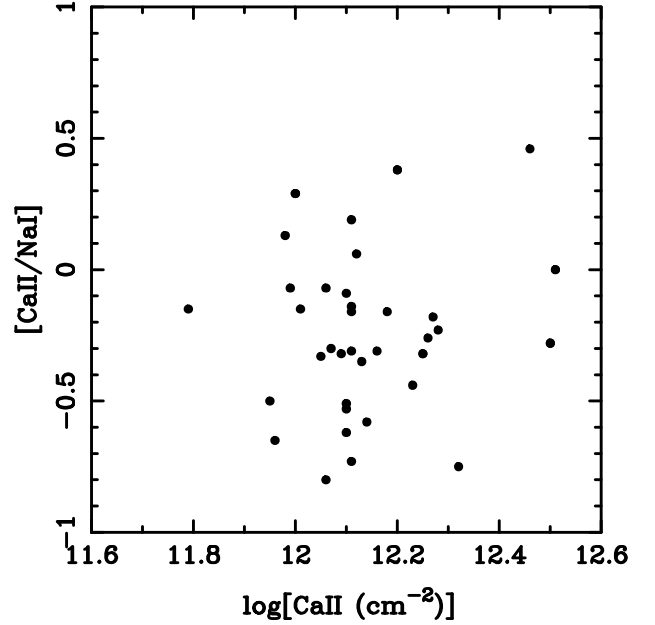}
\caption[]{[Ca\,{\sc ii} K/Na\,{\sc i} D] ratio plotted against log[Ca\,{\sc ii} K] for low velocity gas for the FEROS and UVES Magellanic sightlines only.}
\label{f_FEROS_CaII_div_NaI_vs_CaII}
\end{figure}

\begin{figure}
\setcounter{figure}{10}
\includegraphics{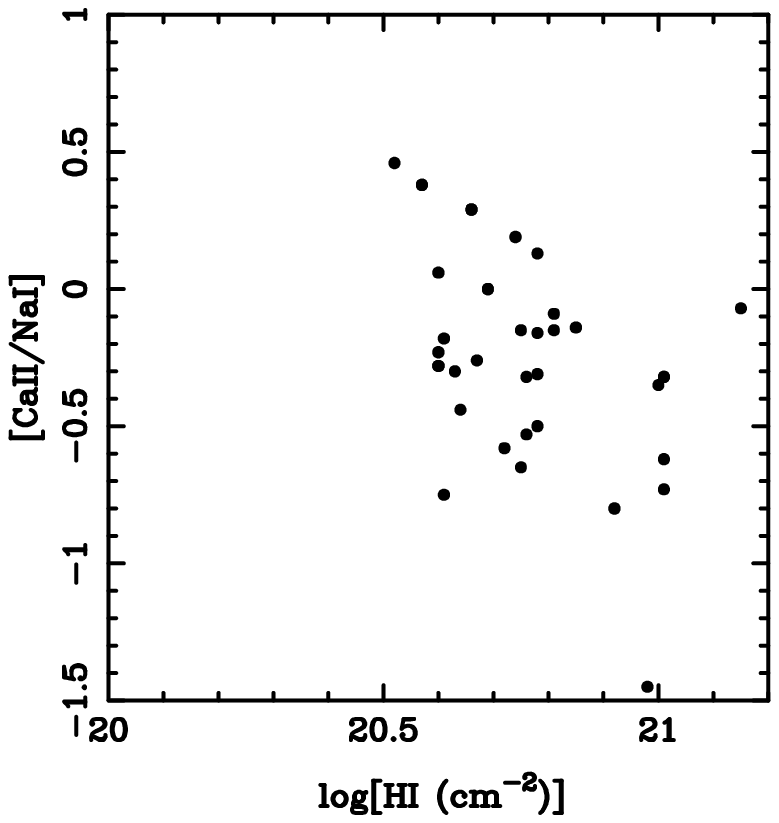}
\caption[]{[Ca\,{\sc ii} K/Na\,{\sc i} D] ratio plotted against log[H\,{\sc i}] for low velocity gas for the FEROS and UVES Magellanic sightlines only.}
\label{f_FEROS_CaII_div_NaI_vs_HI}
\end{figure}

\subsection{The parallax -- column density correlation for Ca\,{\sc ii} and other species}
\label{s_parallax_coldens}

Authors including Beals \& Oke (1953), Megier et al. (2005, 2009) and Welsh et al. (2010) find a slowly-increasing Ca\,{\sc ii} equivalent width 
with increasing distance, although with a large scatter. At distances greatly exceeding 100 pc, Welsh et al. (1997) 
note that the increase in column density is more associated with the number of clouds sampled along a particular 
sight-line as opposed to the actual distance. Figure \ref{f_par_dist}(a) shows the Hipparcos parallax plotted against 
the log of the Ca\,{\sc ii} K column density for all the objects in the current paper for which both quantities are 
available. Additionally, we have used results from Sembach et al. (1993), Hunter et al. (2006) and Smoker et al. (2011) to 
produce a sample of 125 sightlines with parallaxes ranging from $\sim$11 mas down to zero. Superimposed on the plot is the best-fit 
line from Megier et al. (2009) which has the form $\pi$=1/(2.29$\times$10$^{-13}\times N_{\rm CaII}$ + 0.77) (n=262), 
where $N_{\rm CaII}$ is in cm$^{-2}$. At small distances (large values of parallax), the equation from Megier et al. predicts 
larger parallaxes at a given log($N_{\rm CaII}$) than observed in our present dataset, ranging from a $\sim$25 per cent difference 
at 100 pc distance, to $\sim$15 per cent  at 200 pc. This is likely just a reflection of the differing sightlines used in the two datasets, as previously 
observed when results from Megier et al.  (2005) and Megier et al. (2009) are compared. We note that the Local Bubble has dimensions of around $\sim$100 pc 
(e.g. Breitschwerdt et al. 1998, Welsh \& Shelton (2009) and references therein), and hence the column densities in this region are 
frequently very low (e.g. Frisch \& York 1983). Therefore, any analytical formula for the parallax/column density is 
likely to fail in this regime. The best fit that we obtain is;

\begin{equation}
\pi(mas)=1/(2.39 \times 10^{-13} \times N_{\rm CaII} (cm^{-2}) + 0.11),
\end{equation}

which is also displayed on the figure. To further evaluate the relationship, we have plotted in Fig. \ref{f_par_dist}(b)-(h) 
the parallaxes against Ca\,{\sc ii} column densities taken from the compilation of Gudennavar et al. (2012), 
the most extensive 
dataset of such measurements available in the literature. In particular, Fig. \ref{f_par_dist}(b) shows 419 data points with absolute values of Galactic 
latitude less than 10.0$^{\circ}$, with the best-fit lines from Megier et al. (2009) and the current result superimposed. 
Figures EA29 to \ref{f_Parallax_vs_column_density_allspecies} (available online) shows the corresponding plot of parallax against column density 
for the 38 species from Gudennavar et al. (2012). Clear correlations (although with large scatters) are visible in 
Al\,{\sc i} (few data points), Ca\,{\sc ii}, D\,{\sc i}  (few data points), Mn\,{\sc ii}, Na\,{\sc i}, O\,{\sc vi}, P\,{\sc ii} 
and Ti\,{\sc ii}. Of these, only Mn\,{\sc ii} and P\,{\sc ii} are typically the dominant ionisation stage in the warm ISM 
(Sembach et al. 2000). Hence it is likely that other factors, such as dust depletion and inherant clumpiness of the
element, causes much of the observed scatter. In particular, S\,{\sc ii} is both the dominant ionisation stage and typically 
the element is thought to be little depleted onto dust grains in the diffuse ISM. However, the parallax/column density 
relationship for this line shows no reduced scatter compared with other elements. 

Finally, in Fig. \ref{f_caii_par_dist} we plot the Hipparcos parallax against the Ca\,{\sc ii} column density, 
the Hipparcos parallax against the spectroscopic parallax and the spectroscopic parallax against the Ca\,{\sc ii} 
column density. Data are taken from Gudennavar et al. (2012) and we only include stars of spectral type V in the 
comparision. The spectroscopic parallaxes used absolute magnitudes and colours from Schmidt-Kaler (1982) and Wegner (1994), 
from which we estimated the reddening and hence the distance to the stars in question. Straight line 
fits between 100 pc and 1000 pc for CaII/Hipparcos, Spectroscopic/Hipparcos and CaII/Spectroscopic result 
in scatters in the ordinate of 0.67, 0.35 and 0.60, respectively. Hence between these distance limits the 
scatter in the parallax vs spectroscopic parallax fit is 0.25 dex or a factor 1.8 smaller than the 
parallax vs Ca\,{\sc ii} column density fit. We have attempted to improve the correlation by only including sightlines
where log(Ca\,{\sc ii}/Na\,{\sc i}) exceeds 0.5, to exclude instances 
with cooler gas present in which 
the Ca\,{\sc ii} is locked up in dust grains. However, the correlation between parallax and Ca\,{\sc ii} is not 
significantly improved when using this subsample. 


\begin{figure*}
\setcounter{figure}{11}
\includegraphics[height=20cm]{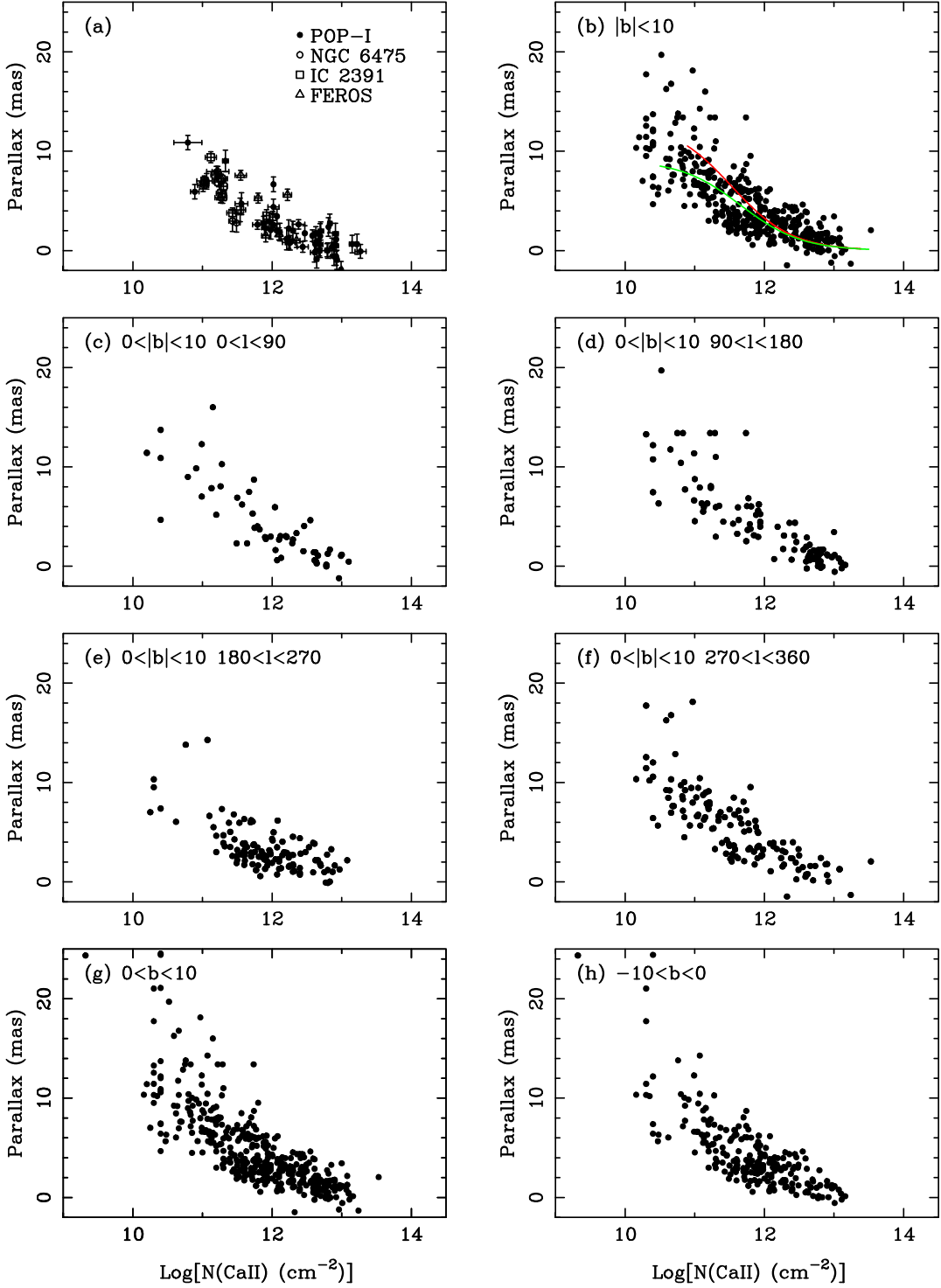}
\caption[]{(a)  Hipparcos parallax against log$N$(Ca\,{\sc ii} K) for stars in the current Galactic FEROS sample plus objects taken from the 
literature. The red line is the best-fit from Megier et al. (2009). The green line is the best fit derived from the current data 
to the function $\pi$=1/(A$\times N$+b) where A=2.39$\times$10$^{-13}$ and b=0.11. (b)-(h) Corresponding plots with data taken from
the compilation of Gudennavar et al. (2012) and a range of Galactic latitude and longitudes.}
\label{f_par_dist}
\end{figure*}

\begin{figure*}
\setcounter{figure}{12}
\includegraphics[]{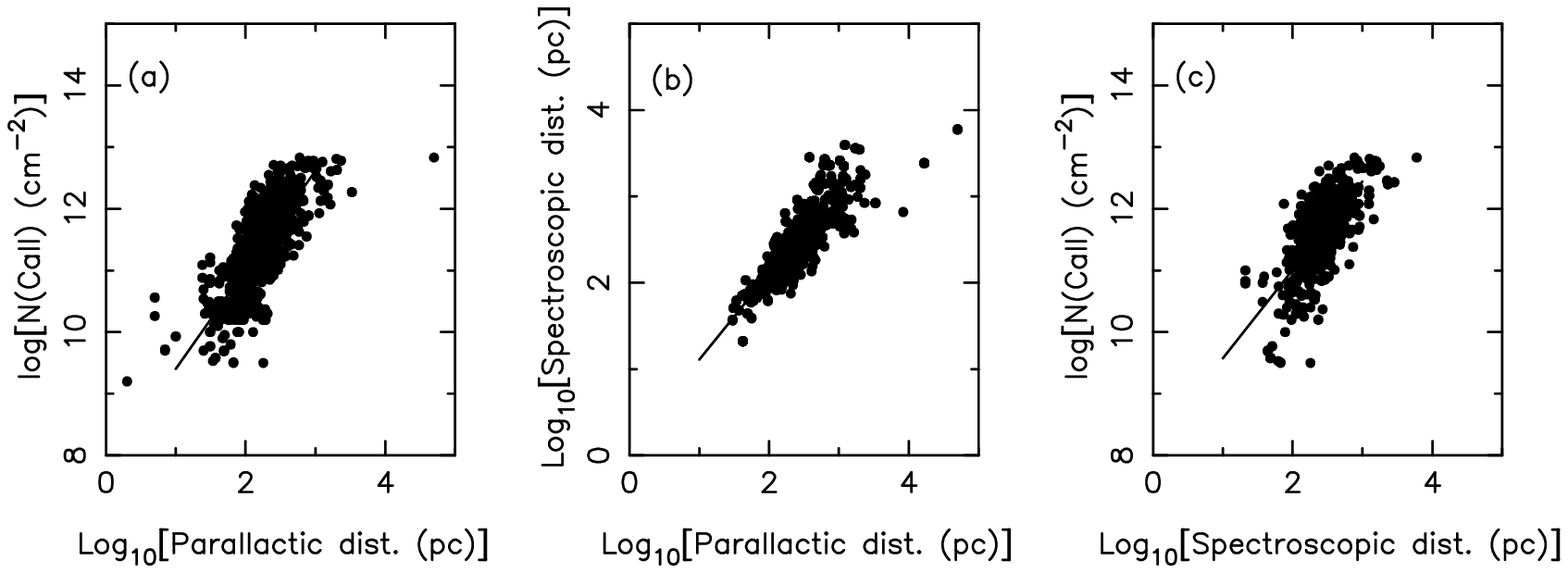}
\caption[]{(a) Log(Ca\,{\sc ii} cm$^{-2}$) against Log(paralactic distance pc). 
(b) Log(Spectroscopic parallax distance pc) against Log(paralactic distance pc) 
(c)  Log(Ca\,{\sc ii} cm$^{-2}$) against Log(Spectroscopic parallax distance pc). 
Parallaxes and spectral types are from {\sc simbad}. Only stars of spectral type 
V are included in the plot. Ca\,{\sc ii} column densities are taken from 
Gudennavar et al. (2012).}
\label{f_caii_par_dist}
\end{figure*}

\section{Summary and suggestions for future work}
\label{s_summary}

We have described the use of FLAMES, FEROS and UVES archive data towards field stars and open clusters in the Milky Way and 
Magellanic Clouds to obtain information on the variability of Ca\,{\sc ii} in the Galactic interstellar medium 
and its possible use as a distance indicator. We find that towards 4 Magellanic open clusters 
%
%
the maximum variation observed is between a factor of $\sim$1.8 and 3 in equivalent width or $\sim$0.3--0.5 dex 
in column density in the optically thin approximation over fields of size $\sim$0.05--6 pc. These observations can 
be explained by a simple model of the ISM presented in van Loon et al. (2009) although likely other functional 
forms of the ISM would also match the observations. Using archive observations and results from the literature we 
derive a parallax -- column density relationship for Milky Way gas in Ca\,{\sc ii} of 
$\pi$(mas)=1/(2.39$\times$10$^{-13}\times N_{\rm CaII}$(cm$^{-2}$ + 0.11) that predicts parallaxes to within 15 percent 
of the Megier et al. (2009) values for distances of 200 pc. 

A future paper will use the FLAMES HR4 grating setting to study the variation in the Ca\,{\sc i} and CH$^{+}$ lines at 4226\AA\ 
and 4232\AA, respectively, for the sample of clusters discussed here, to determine the variation in neutral and 
molecular column density. New observations using the Na\,{\sc i} D line could probe the variation in the Routley-Spitzer effect 
on small scale and how it changes with reddening. 

\begin{acknowledgements}

This paper makes use of data taken from the Archive of the European Southern Observatory. This research has made use of the {\sc simbad} database,
operated at CDS, Strasbourg, France. JVS thanks Queen's University Belfast for financial support under the 
visiting scientist scheme and to ESO for financial support 
under the Director General's Discressionary fund. We would like to thank an anonymous referee for their careful reading of the
original manuscript.

\end{acknowledgements}

  \end{document}